# Linear and nonlinear electrodynamic response of bulk $CaC_6$ in the microwave regime

A. Andreone*, G. Cifariello, E. Di Gennaro

*CNISM and Department of Physics, University of Naples "Federico II", 80125 Naples, Italy*

G. Lamura

*CNR-INFM Coherentia and Department of Physics, University of Naples "Federico II", 80125 Naples, Italy*

N. Emery, C. Hérold, J. F. Marêché, P. Lagrange

*Laboratoire de Chimie du Solide Minéral-UMR 7555, Université Henri Poincaré Nancy I, B.P. 239, 54506 Vandœuvre-lès-Nancy Cedex, France*

**Abstract.** The linear and nonlinear response to a microwave electromagnetic field of two c-axis oriented polycrystalline samples of the newly discovered superconductor $CaC_6$ ($T_C \approx 11.5$ K) is studied in the superconducting state down to 2 K. The surface resistance $R_S$ and the third order intermodulation distortion, arising from a two-tone excitation, have been measured as a function of temperature and microwave circulating power. Experiments are carried out using a dielectrically loaded copper cavity operating at 7 GHz in a "hot finger" configuration. The results confirm recent experimental findings that $CaC_6$ behaves as a weakly-coupled, fully gapped, superconductor. The weak power dependence of $R_S$ encourages a further investigation of this novel superconductor as a possible alternative to Nb in specific microwave applications.

*PACS*: 74.25.Nf, 74.70.Ad, 71.20.Tx

One of the main priorities following the discovery of a novel superconductor, outside of a full comprehension of the microscopic mechanisms involved in the condensate state, is the study of its electrodynamic properties, in order to understand whether the new material can be useful for practical applications. Amongst other parameters, the response of the superconductor to an e.m. field at high frequencies is an important test to determine its possible use in resonant cavities for particle accelerators or in passive devices for mobile and satellite communications.

The new superconductor $CaC_6$, setting a record $T_C$ of 11.5 K in the family of graphite intercalated compounds (GIC) [1, 2], is potentially able to replace niobium for the internal coating of accelerating cavities, or in other niche applications (like extremely stable oscillators or highly selective filters) where performance requirements are more stringent than cryogenic issues.

There are several reasons for considering graphite intercalated with Ca a candidate alternative to Nb:

- there is a mounting experimental evidence, from different spectroscopic tools like London penetration depth [3], tunnelling [4], and specific heat [5], that the superconducting order parameter of $CaC_6$ follows a conventional isotropic s-wave symmetry like Nb does, and presents a ratio *2Δ(0)/k_BT_C* – where *Δ(0)* is the zero temperature energy gap – close to the BCS prediction;
- the normal state transport properties are as good as or even better than Nb [6, 7];
- even if only the synthesis of bulk samples has been reported so far, the simple binary nature of $CaC_6$ allows to predict that clean samples in thin film form could be available quite soon. To corroborate the optimism of this assertion, the successful deposition of thin films of graphene, one of the two constituent layers of $CaC_6$ [2], appeared in recent literature [8].

We report here a study of the linear and non linear microwave properties of bulk samples of this new superconductor, synthesized from highly oriented pyrolytic graphite [9]. The results are then compared with measurements performed on Nb.

Data have been taken on two platelike c-axis oriented polycrystals having a roughly squared shape of maximum size 2.5x2.5 mm$^2$ and thickness of 0.1 mm. Because CaC$_6$ is highly reactive when exposed to oxygen, the samples were accurately cleaved in an inert atmosphere before each run.

For measuring the electrodynamic response of the samples under test, we used an open-ended dielectric single-crystal sapphire puck resonator operating at the resonant frequency of 7 GHz in a "hot finger" configuration. The resonator enclosure is made of oxygen-free high conductivity copper, whereas the sample holder is a low loss sapphire rod, placed at the centre of the cavity in close proximity to the puck dielectric crystal. The cavity is excited with a transverse electric TE$_{011}$ mode, which induces a-b plane screening currents in the sample. By using a micrometer screw, the position of the sample placed on the sapphire rod, and therefore the puck-to-sample distance, can be changed, in order to get the maximum sensitivity. The apparatus allows any sample between 0.1 and 100 mm$^2$ to be placed in the cavity. The capability to move the sample and make the cavity sensitivity variable allows to measure surface resistance values ranging between Ohms and hundreds of μOhms, depending on the sample dimensions. Unfortunately, the actual apparatus is not free of systematic errors, presently making difficult to extract the sample surface reactance (and the penetration depth) from the frequency shift data.

The experimental data are taken using two different circuital configurations:

*i)* a single-tone mode, where the cavity is excited at its resonance by a sweeping field, and the surface resistance $R_S$ of the sample under test is extracted as a function of temperature *T* and microwave surface field $H_{r.f.}$, using a standard perturbation method. Measurements are performed via a vectorial network analyzer. Of course, in order to ensure that the (virtually) zero field surface resistance is being extracted, care must be taken to keep power at a suitably low level. To study the behavior of $R_S(H_{r.f.})$, a linear amplifier is used, allowing to raise the power circulating in the cavity up to some kW;

*ii)* a two-tone mode, to study the third order intermodulation distortion (IMD) generated in the CaC$_6$ samples. Briefly, two pure frequencies f$_1$ and f$_2$ (> f$_1$) with equal amplitudes are generated by two phase-locked synthesizers. The frequencies are separated symmetrically around the centre frequency and spaced by 10 KHz, so that both $f_1$ and $f_2$ are well within the 3 dB bandwidth of the resonator. The two signals are combined and applied to the resonant cavity where the sample under test has been inserted, and the output signals (the two main tones and the two third order IMD's at $2f_1-f_2$ and $2f_2-f_1$) are then monitored as a function of the input power by using a spectrum analyser. Further details can be found in [10, 11].

The microwave surface resistance $R_S(T)$ of bulk $CaC_6$ is shown in Fig. 1, showing the expected sharp transition at $T_C$ and a rapid drop as T decreases. At temperature well below $T_C$, $R_S$ tends to saturate, reaching a high residual value more or less below 3 K. In spite of the good quality of the sample, the data cannot be consistently fitted in the overall temperature range within a BCS framework, since extrinsic surface effects are dominant at the lowest temperatures. These residual losses come very likely from the presence of calcium oxides and hydroxides on the $CaC_6$ surface, in spite of the care taken in keeping the sample in a controlled environment during each run.

For comparison, in the same figure we show data taken on a high quality Nb commercial sample (Goodfellow) having a similar size as $CaC_6$. In this last case, $R_S$ values are well below the limit of sensitivity (evidenced by the hatched region in the graph) already at 6 K.

For *T* less than $T_c/2$, in conventional superconductors the surface resistance behavior can be phenomenologically described using the standard BCS exponential dependence, after subtracting to the data a residual term $R_{res}$ related to the extrinsic losses [12]. This is done in the inset of figure 1, where the quantity $R_S - R_{res}$ is displayed as a function of $T_C/T$ at low temperatures, and $R_{res} = R_S(T_{min})$ is chosen. Once the residual value is subtracted, the low temperature data does show an exponential temperature dependence, according to $R_S \propto exp-(\Delta(0)/k_BT)$. A straight line behavior is clearly seen at low *T*, that is an unambiguous and direct signature of the superconducting gap. A fit has been performed for $T_c/T$ ranging between 2 and 4 only (that is, approximately from 6 to 3 K), because of the large scattering of data at the lowest temperatures, giving a strong coupling ratio

$2\Delta(0)/k_BT_C = (3.6 \pm 0.5)$. The resulting gap $\Delta(0) = (1.7 \pm 0.3)$ meV is compatible with the results obtained from penetration depth (1.79 meV) [3] and STM (1.6 meV) measurements [4].

Supposing that the electrodynamic response of $CaC_6$ in the normal state is local - consistently with the experimental observation that the screening response in the superconducting state lies in the dirty limit [3] - we can calculate the microwave resistivity $\rho_n$ using the formula for the classical skin depth: ($R_S = (\pi\mu_0 f\rho_n)^{0.5}$), where $\mu_0$ is the vacuum permeability and $f$ is the resonance frequency. The residual term $\rho_{n0}$, defined as the onset of the superconducting transition, is close to 5 $\mu\Omega\cdot$cm, consistent with values reported in literature for the dc resistivity measured on samples from the same source [7]. The low temperature (T << $\Theta_D \approx$ 600 K [5], where $\Theta_D$ is the Debye temperature) dependence of $\rho_n$ is shown in fig. 2 to follow a power law $T^n$ (continuous line), with the exponent $n$ ranging between 2 and 3 (in the graph n = 2.5). Assuming that electron-phonon scattering is the dominant contribution determining the resistivity of $CaC_6$, the observed power law behaviour shows some deviation as to the universal ($\rho_{n0}$, $\lambda$) graph of all metals [13], $\lambda$ being the coupling constant. According to this phenomenological plot, there are two reasonably well separated regions of $\rho_{e-ph}(T)$ behavior at low temperature, $T^{3-5}$ and $T^2$ respectively, depending on the values of ($\rho_{n0}$, $\lambda$). The n = 3 to n = 2 transition usually takes place increasing both the electron-phonon coupling ($\lambda$) and the disorder ($\rho_{n0}$). Metals with $\lambda \leq 0.9$ should never enter nor being close to the $T^2$ region, independent of the amount of disorder. The observed discrepancy can be an indication that the strong coupling constant of $CaC_6$ might be a bit larger than previously calculated ($\lambda \approx 0.85$) [14, 5].

In fig. 3 the behaviour of $R_S$ as a function of the microwave surface field $H_{rf}$ is reported at six different temperatures. The $R_S(H_{rf})$ dependence is extremely flat for temperatures as high as 8 K, which is quite encouraging, taking into account the polycrystalline nature of the samples. Above 8 K, the dependence of $R_S$ on the field is more pronounced, with a steeper slope but with no evidence of a sharp transition related to some threshold field [10].

One must note, however, that measurements carried out at 4 K on the Nb bulk sample of similar size used for comparison show no dependence at all, up to the maximum available circulating power $P_{circ}$ (above 1 KW).

In the extremely low power region - where the surface resistance of $CaC_6$ shows no detectable dependence on the microwave field at all - the IMD response follows the expected cubic dependence on $P_{circ}$ [15]. According to the theory first set by Dahm and Scalapino [16], the non linear response of a superconductor measured by $P_{IMD}$ (the output power due to the third order IMD) can be used as a very sensitive probe of the superconducting gap function symmetry, by means of the nonlinear parameter $b(T)$. This can be done exploiting the relation between $P_{IMD}$ and $b^2$ [17]. Taking into account the same rescale method used successfully in ref. [11], we have extracted the nonlinear parameter $b^2$ as a function of the temperature for a fixed value of the circulating power $P_{circ}$. The results of this procedure are shown in figure 4. Experimental data for $CaC_6$ follows the curve (solid line) predicted by the theory [18] for an isotropic, single-gapped superconductor, confirming previous experimental findings.

In summary, we have presented a study of the electrodynamic response of bulk $CaC_6$ samples at 7 GHz as a function of temperature and microwave power. In spite of the fact that the surface properties of this GIC in the microwave region are far from being optimized, the temperature dependence of its surface resistance, cleared of extrinsic effects, can be well explained within the conventional BCS theory. A single-valued and finite gap can be extracted from the data, consistent with recent measurements of the superfluid density performed on samples from the same source. In a similar way, data from IMD measurements indicate that the observed nonlinearity has an intrinsic origin and its temperature dependence follows the behavior expected for an s-wave superconductor. Finally, the power dependence of $R_S$ is encouraging, in spite of the polycrystalline nature of the samples under test, making $CaC_6$ an attractive alternative to Nb in some microwave applications and deserving further experimental studies. Certainly, the availability of samples in thin film form will call for new experiments.

# References

[*] Corresponding author. Tel.: +390817682547; fax: +390812391821; e-mail: **andreone@unina.it**

[1]  T. E. Weller *et al*., Nature Phys. **1**, 39 (2005).

[2]  N. Emery *et al.*, Phys. Rev. Lett. **95**, 087003 (2005).

[3]  G. Lamura *et al.*, Phys. Rev. Lett. **96**, 107008 (2006).

[4]  N. Bergeal *et al.*, Phys. Rev. Lett. **97**, 077003 (2006).

[5]  J. S. Kim *et al.*, Phys. Rev. Lett. **96**, 217002 (2006).

[6]  M. S. Dresselhaus and G. Dresselhaus, Adv. Phys. **51**, 1 (2002).

[7]  A. Gauzzi *et al.*, cond-mat/0604208 (2006).

[8]  T. Ohta *et al.*, Science **313**, 951 (2006).

[9]  N. Emery *et al*., J. Solid State Chem. **178**, 2947(2005).

[10] G. Lamura *et al.*, Appl. Phys. Lett. **82**, 4525 (2003).

[11] G. Cifariello *et al.*, Appl. Phys. Lett. **88**, 142510 (2006).

[12] M. Tinkham, *Introduction to Superconductivity*, McGraw-Hill, New York (1996).

[13] M. Gurvitch, Phys. Rev. Lett. **56**, 647 (1986).

[14] M. Calandra and F. Mauri, Phys. Rev. Lett. **95**, 237002 (2005).

[15] R. Monaco, A. Andreone, and F. Palomba, J. Appl. Phys. **88**, 2898 (2000).

[16] T. Dahm and D. J. Scalapino, Appl. Phys. Lett. **69**, 4248 (1996).

[17] T. Dahm and D. J. Scalapino, J. Appl. Phys. **81**, 2002 (1996).

[18] E. J. Nicol, J. P. Carbotte, and D. J. Scalapino, Phys. Rev. B **73**, 014521 (2006).

**Figure captions**

**Fig. 1**. $R_S$ vs $T$ for CaC$_6$ sample #1 (○). Data obtained for a Nb bulk sample are shown for comparison(▼). The hatched region indicates the sensitivity limit of the experimental setup. In the inset: ($R_S$-$R_{res}$) vs $T_c/T$ is plotted below $T_c/2$ on a semi-log scale for the same sample. The continuous line represents the exponential behavior predicted by the BCS theory with a strong coupling ratio $2\Delta(0)/k_BT_C = (3.6 \pm 0.5)$.

**Fig. 2.** Microwave resistivity in the normal state for T<<$\Theta_D$ (CaC$_6$ sample #1). The dotted line is a numerical fit for the expression $\rho = a + bT^n$. In this graph n = 2.5.

**Fig. 3.** $R_S(H_{rf})$ is shown at various temperatures for CaC$_6$ sample #1.

**Fig. 4.** The nonlinear parameter $b^2$ (proportional to the IMD power) vs the reduced temperature $t = T/T_c$ for CaC$_6$ sample#2. Continuous line represents the temperature behavior for the non linear coefficient expected in the case of the standard *s*-wave model.

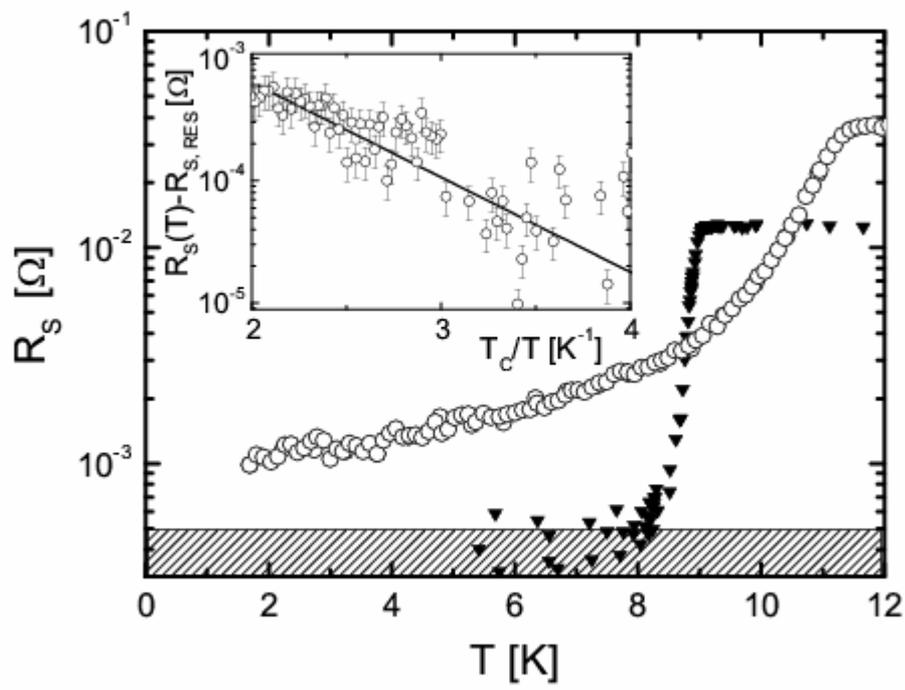

**Fig. 1**

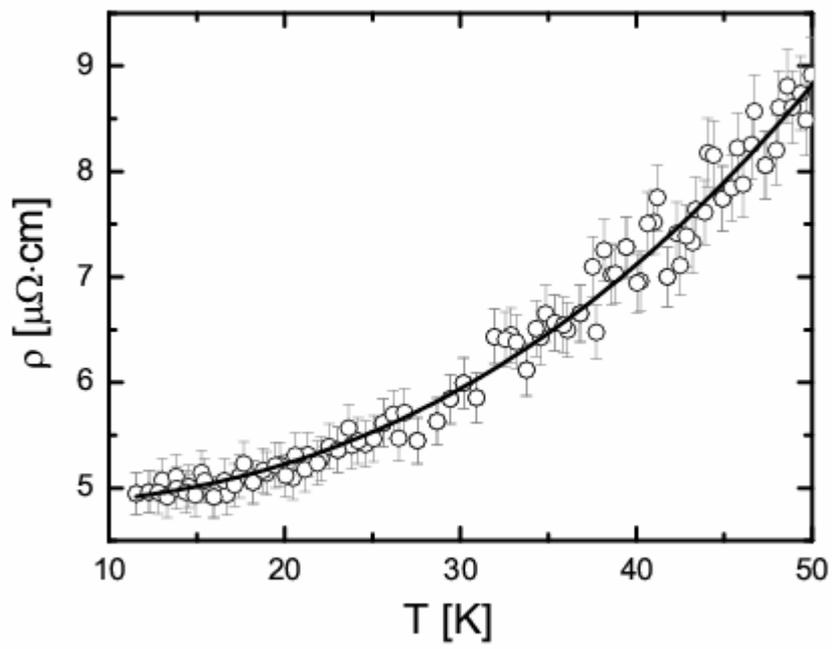

Fig. 2

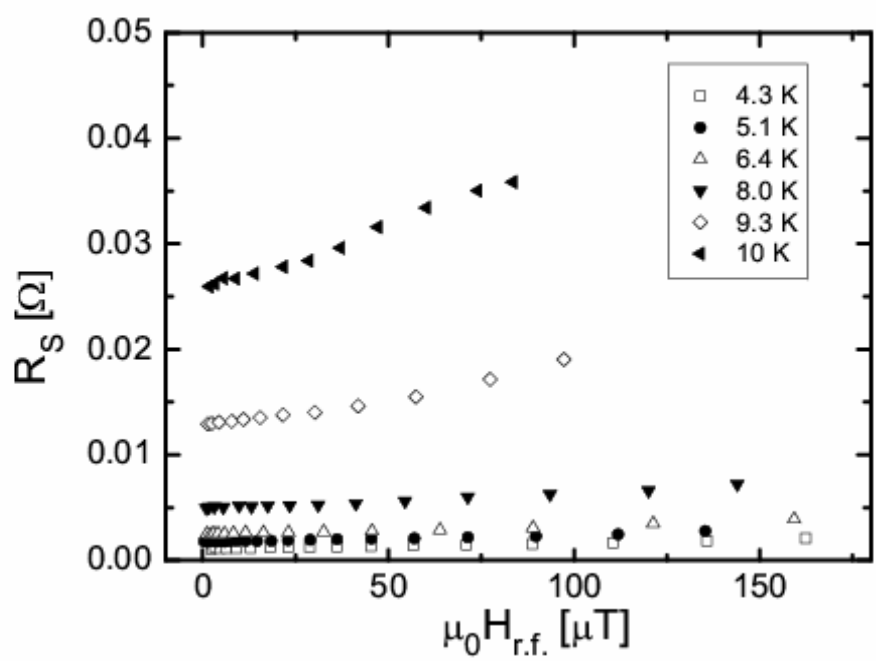

**Fig. 3**

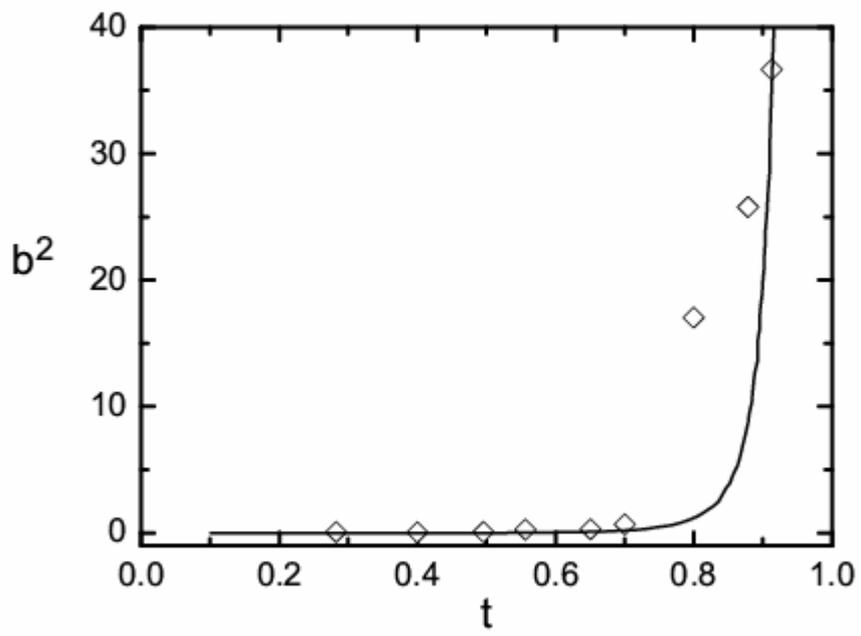

**Fig. 4**